\documentclass[preprint,aps, showpacs]{revtex4}
\usepackage{graphicx}% Include figure files
\begin{document}

%\preprint{}

\title{{\large Connection between Second Class Currents and the $\Delta
N\gamma$ Form Factors $%
G_{M}^{\ast }(q^{2})$ and $G_{E}^{\ast }(q^{2})$}}
\author{Milton Dean Slaughter}
\address{Department of Physics, University of New Orleans, New Orleans, LA 70148}
\date{December 2004}
 \email{E-Mail: mslaught@uno.edu}

\begin{abstract}
An interesting connection between the nucleon weak axial-vector
second class current form factor $g_{T}(q^{2})$ present in the
matrix
element $\left\langle {p}%
\right\vert A_{\pi ^{+}}^{\mu }\left\vert {n }\right\rangle $ and
the $\Delta
N\gamma$ form factors $%
G_{M}^{\ast }(q^{2})$ and $G_{E}^{\ast }(q^{2})$ is derived. Using
a nonperturbative, relativistic sum rule approach in the infinite
momentum
frame, $%
G_{M}^{\ast }(q^{2})$ and $G_{E}^{\ast }(q^{2})$ are calculated in
terms of $g_{T}(q^{2})$ and the well-known nucleon isovector Sachs
form factor $G_{M}^{V}$ as input with no additional model
parameters. Reasonable agreement with the data for $%
G_{M}^{\ast }(q^{2})$ may be achieved with a non-zero
$g_{T}(q^{2})$ too large to be accommodated in the Standard Model.
  We surmise that it is plausible that
second class current-associated pion cloud effects are playing a
significant role in pion electroproduction processes and perhaps
must be taken into account in those methodologies which utilize
effective Lagrangians.
\end{abstract}


\maketitle

%%%next statement makes entire doc Large
%%%Don't forget matching brace at doc end

The study of the nucleon weak axial-vector second class current
(SCC)\protect\cite{weinberg} form factor $g_{T}(q^{2})$ present in
the matrix
element $\left\langle {p}%
\right\vert A_{\pi ^{+}}^{\mu }\left\vert {n }\right\rangle $ and
 the $\Delta N\gamma$ form factors $G_{M}^{\ast
}(q^{2})$ and $G_{E}^{\ast }(q^{2})$ has engendered much
experimental and theoretical research for several
decades\protect\cite{holstein71,kubodera,monsay,jonesscadron,
ahrens88,stoler93,shiomi96,
tminamisono1998,wilkinson2000nim, kminamisono2000,
gardner2001,abele2002,kuzminaug2004}. In a perfect world with
unbroken $SU_{F}(N)$ flavor symmetry, one
expects that $G_{E}^{\ast }(q^{2})=0$ and that $%
G_{M}^{\ast }(q^{2})$ would exhibit the same $q^{2}$ behavior as
does the Sachs nucleon form factor $G_{M}$.  Instead, one
unexpectedly finds that, experimentally, $G_{M}^{\ast }$ decreases
faster as a function of $Q^{2}\equiv -q^{2}$ than does $G_{M}$ and
furthermore, that not only does the ratio $-G_{E}^{\ast
}/G_{M}^{\ast }\neq 0$ but indeed,
possesses a complicated
behavior as a function of $Q^{2}$.  In this Letter, we suggest
that the inclusion of SCC effects normally neglected in most
experimental and theoretical studies in general and in pion
photoproduction and electroproduction processes in particular, may
aid in our understanding of the basic underlying connecting
physics principles responsible for such processes.

We begin by using a nonperturbative, relativistic sum rule
approach in the infinite momentum
frame where $%
G_{M}^{\ast }(q^{2})$ and $G_{E}^{\ast }(q^{2})$ are calculated in
terms of $g_{T}(q^{2})$ and the well-known nucleon isovector Sachs
form factor $G_{M}^{V}$ as input with no additional model
parameters{\protect\cite{miltnpasub2003}. With regard to shape and
data, reasonable agreement
 for $%
G_{M}^{\ast }(q^{2})$ may be achieved with a non-zero
$g_{T}(q^{2})$ probably too large to be accommodated in the
Standard Model (SM).  We surmise that it is plausible that
SCC-associated pion cloud effects are playing a significant role
in pion photoproduction and electroproduction processes and
perhaps must be taken into account in those methodologies which
utilize effective Lagrangians.

The most general form for the matrix element of the weak current
between the proton and the neutron is given by

\begin{eqnarray}
& &(2\pi)^{3} \sqrt{\frac{E_{p_{1}}E_{p_{2}}}{mm_{n}}}
\left\langle P(p_{2})\right| A_{\pi^{+}}^{\mu}(0)\left|
N(p_{1})\right\rangle
\nonumber \\ &=&\bar{u}_{P}%
(p_{2})\left[
\{g_{A}({\tilde{q}}^2)\gamma^{\mu}+ig_{T}({\tilde{q}}^2
)\sigma^{\mu\nu}{\tilde{q}}_{\nu}+ig_{P}({\tilde{q}}^2)
{\tilde{q}}^{\mu}\}\gamma_{5}
\right]
u_{N}(p_{1}). \label{Eq1}%
\end{eqnarray}

The 4-momentum transfer ${\tilde{q}}^2=p_{1}-p_{2}$.  $g_{A}$,
$g_{T}$, and $g_{P}$ are the axial-vector, induced pseudotensor,
and induced pseudoscalar form factors respectively. With respect
to transformations under G-parity, $g_{A}$ and $g_{P}$ are
represented by first-class currents (FCC) while $g_{T}$ is
represented by SCC. In the SM, a non-zero $g_{T}$ can only arise
due to quark mass and charge differences and thus the ratio
$g_{T}/g_{A}$is thought to be very small or identically zero.
  In Ref.{\protect\cite{miltnpasub2003}---{\em assuming no SCC effects}---the $\Delta
N\gamma $ transition form
   factors $G_{M}^{\ast }(q^{2})$ and $%
G_{E}^{\ast }(q^{2})$ were calculated in terms of well-known
nucleon isovector Sachs form factor parametrized by $G_{M}^{V}({\tilde{q}}^{2})=\frac{1}{%
2}(\mu _{p}-\mu _{n})G_{\mbox{dipole}}({\tilde{q}}^{2})$, with
$G_{\mbox{dipole}}({\tilde{q}}^{2})\equiv \lbrack
1-{\tilde{q}}^{2}/0.71 \ GeV^2/c^2]^{-2}$, where $\mu _{p}$ and
$\mu _{n}$ are the proton and neutron magnetic moments
respectively.  On the other hand, {\em if one now allows for SCC
effects}, one obtains:
\begin{eqnarray}
G_{M}^{\ast }(q^{2})& = &
\left(3c_{2}(q^2)G_{M}^{V}({\tilde{q}}^{2}_{+})+c_{1}(q^2)c_{3}(q^2)
\sqrt{
{\tilde{Q}}^{+}_{+}{\tilde{Q}}^{-}_{-}}G_{M}^{V}({\tilde{q}}^{2}_{-})\right)/
\left((3+c_{1}(q^2))\sqrt{{\tilde{Q}}^{+}_{+}}\right), \label{Eq2} \\
G_{E}^{\ast }(q^{2})& = &
\left(c_{2}(q^2)G_{M}^{V}({\tilde{q}}^{2}_{+})-c_{3}(q^2)\sqrt{
{\tilde{Q}}^{+}_{+}{\tilde{Q}}^{-}_{-}}G_{M}^{V}({\tilde{q}}^{2}_{-})\right)/
\left((3+c_{1}(q^2))\sqrt{{\tilde{Q}}^{+}_{+}}\right), \label{Eq3}
\end{eqnarray}

\begin{eqnarray}
c_{1}(q^2) & = & \frac{m^{*}(4m-m^{*})+m^{2}-q^{2}}{%
(m^{*2}+m^{2}-q^{2})},  \label{Eq4} \\
c_{2}(q^2)& = & \frac{5\sqrt{3}}{3}
\left[
-\left(\frac{m^{*}m}{m^{*}+m}\right)
\left(\frac{{\tilde{Q}}^{+}_{+}\sqrt{{\tilde{Q}}^{-}_{+}}}
{(m^{*2}+m^{2}-q^{2})G_{M}^{V}({\tilde{q}}^{2}_{+})}
\right) \label{Eq5}
\left(\frac{g_{T}({\tilde{q}}^{2}_{+})}{g_{A}(0)\sqrt{4\pi\alpha}}\right) \right.  \\
\nonumber & + &
\left.
\frac{2m^{*}m^{2}\sqrt{Q^{+}}}{(m^{*}+m)(m^{*2}+m^{2}-q^{2})}  \right],
 \\
c_{3}(q^2) & = &  \frac{m} { (m^{*}+m)\sqrt{{\tilde{Q}}^{-}_{+}}},
\label{Eq6}
\end{eqnarray}

\noindent where $\alpha \equiv$ fine-structure constant, $q\equiv p^{\ast }-p$, $p^{\ast }=(%
p^{\ast 0},\vec{t})$ and $p=(p^{0},\vec{s%
})$ are the four-momenta of the $\Delta ^{+}$ and nucleon
respectively, $m^{\ast }= \Delta ^{+} $ mass, $m=$ proton mass
$\approx$ neutron mass $=m_{n}$,
 $\widetilde{%
q}=\widetilde{p}^{\ast }-\widetilde{p}$, $p_{1}=\widetilde{p}^{\ast }=(%
\widetilde{p}^{\ast 0},\vec{t})$ and
$p_{2}=\widetilde{p}=(\widetilde{p}^{0},\vec{s%
})$,
 $Q^{\pm }\equiv (m^{\ast }\pm m)^{2}-q^{2} $,\linebreak
${\tilde{q}}^{2}_{\pm}=[(m^{\ast 2}+m^{2})q^{2}+ (m^{\ast
2}-m^{2})(\pm\sqrt{Q^{+}Q^{-}}-(m^{\ast 2}-m^{2}))]/(2m^{\ast
2})$,
 ${\tilde{Q}}^{+}_{\pm}\equiv
4m^2-{\tilde{q}}^{2}_{\pm}$, and ${\tilde{Q}}^{-}_{\pm}\equiv
-{\tilde{q}}^{2}_{\pm}$.  Note that ${\tilde{q}}^{2}_{+}(q^2=0)=0$
 and also that $g_{P}$ does not contribute to $G_{M}^{\ast }$ and $G_{E}^{\ast
}$ in Eqs.(\ref{Eq2}) and (\ref{Eq3}). While very little is known
about $g_{T}$, it is traditional
 to model $g_{T}$ as a dipole
similar to $g_{A}$ and $G_{M}$: $g_{T}( {\tilde{q}}^{2}_{+})
)=g_{T}(0)\lbrack 1-{\tilde{q}}^{2}_{+}/m_{T}^{2}]^{-2}$, where
$m_{T}$ is a ``pseudotensor mass" analogous to the axial-vector
mass $m_{A}\approx 1.2 \, GeV/c^2$.  However, we note, that at
present, no theoretical justification for a specific form of
$g_{T}$ is known. Indeed, we find that an exponential form such as
$g_{T}({\tilde{q}}^{2}_{+}) )=g_{T}(0)\lbrack
exp({\tilde{q}}^{2}_{+}/m_{T}^{2})]$ may also suffice.  We also
note that $G_{M}^{\ast }(q^{2})$ as defined above is related to
another widely used phenomenological form factor $G_{M}^{\ast Ash
}(q^{2})$ \protect\cite{ash} by $G_{M}^{\ast }(q^{2})=G_{M}^{\ast
Ash }(q^{2})\sqrt{1-q^{2}/(m^{\ast }+m)^{2}}$
\protect\cite{jonesscadron}.

From Eqs. (\ref{Eq2}--\ref{Eq6}) and given a specific form for
$g_{T}$, one may calculate $G_{M}^{\ast }$ and $G_{E}^{\ast }$. In
Fig. 1, we present our results for $G_{M}^{\ast Ash  }(Q^{2})$
normalized relative to $3 \,G_{\mbox{dipole}}(Q^{2})$ along with
experimental data.  In Fig. 2, we
present our results for
$G_{E}^{\ast }(Q^{2})$ in terms of the ratio $R_{EM}\equiv
-G_{E}^{\ast }/G_{M}^{\ast }$.  As is evident, an adequate fit may
be achieved by assuming a non-zero $g_{T}$.  Interestingly, an
exponential form for $g_{T}$ suffices as well as a dipole form
when one considers $R_{EM}$ and its behavior in the region $0\leq
Q^{2}\lesssim 1 \; (GeV/c)^{2}$.

In addition to demonstrating the faster than dipole decrease in
$G_{M}^{\ast }$ as a function of $Q^2$---in agreement with
experiment---and the change in sign of $R_{EM}$ as $Q^2$
increases---as indicated by experiment---the curves in Fig. 1 and
Fig. 2 suggest that the small (close to the real photon point)
$Q^2$ behavior  of both $G_{M}^{\ast }$ and $G_{E}^{\ast }$ may be
much more complex than one may have perhaps
anticipated and indeed
may signal the presence of a SCC contribution to basic pion
electroproduction processes.  If this is indeed the case, a
possible explanation could be pion cloud effects associated with
the matrix
element $\left\langle {p}%
\right\vert A_{\pi ^{0}}^{\mu }\left\vert {p }\right\rangle \propto
\left\langle {p}%
\right\vert A_{\pi ^{+}}^{\mu }\left\vert {n }\right\rangle $
which may have
to be explicitly included in dynamical approaches to pion
photoproduction and electroproduction.

The author is grateful to Professor Paul Stoler for providing data
used in this work and for very useful and provocative
communications.
%\bigskip
%\newpage

%%%Next brace belongs to enlarging command at doc beginning

%%Figure caption is next

\begin{figure}[tbp]
\includegraphics[height=5.14in,width=6.5in]{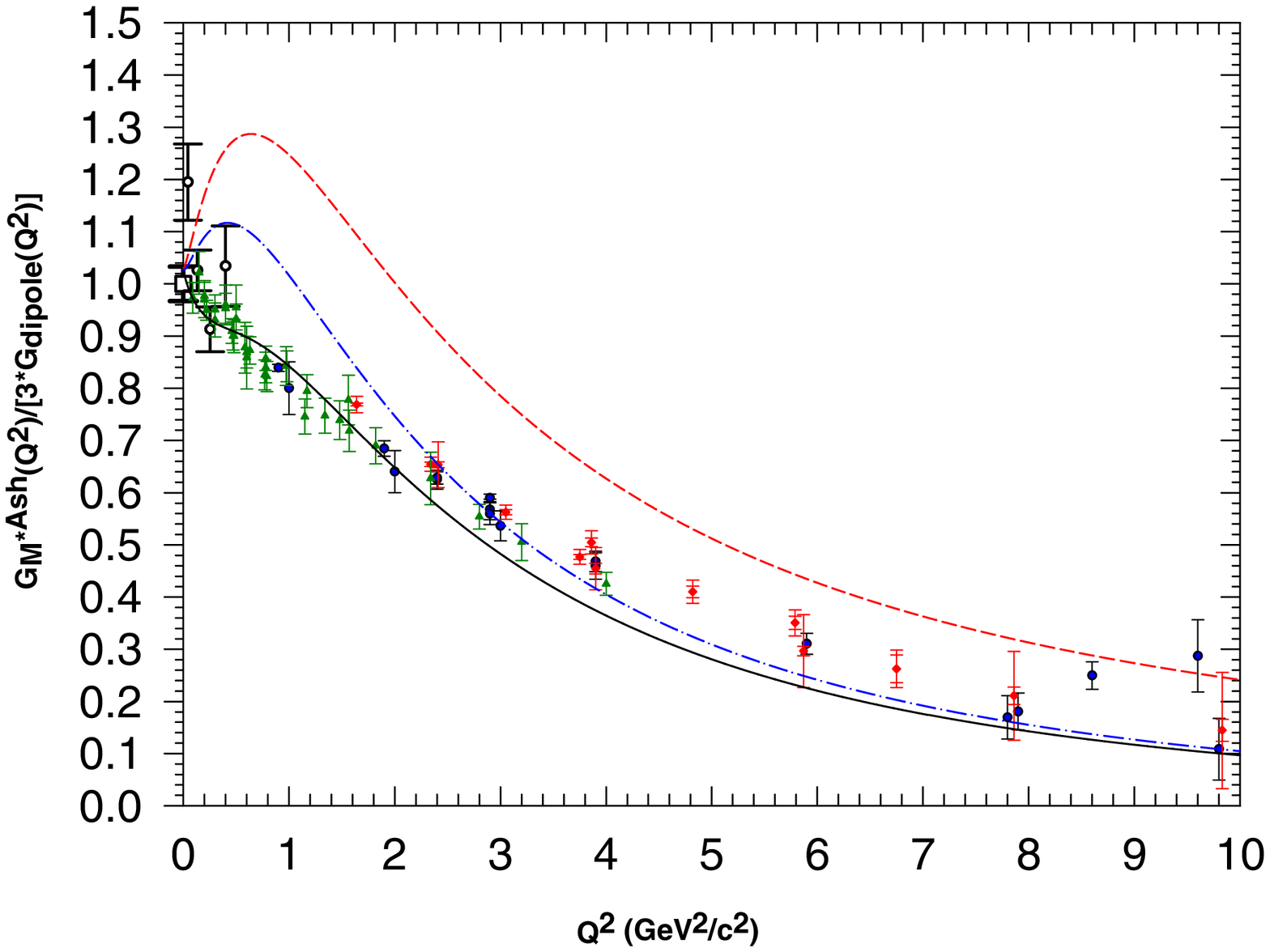}
\caption{$G_{M}^{\ast Ash  }(Q^{2})$ normalized to $3 \,
%G_{\mbox{dipole}}(Q^{2})$,$\, R_{EM}(Q^{2})$, and $R_{SM}(q^{2})$.
G_{\mbox{dipole}}(Q^{2})$.  Theoretically calculated Dashed curve
with $g_{T}=0$ as discussed in the text with $G_{M}^{\ast}(0)/3 \,
G_{\mbox{dipole}}(0)=1.03$;  The Solid curve is a $g_{T}$ dipole
fit to the JLAB $G_{M}^{\ast}$ data of Ref.
\protect\cite{kamalov1} where $g_{T}(0)/g_{A}(0)=1.169 \;
c^2/GeV$, $m_{T}=0.534 \; GeV/c^2$ is obtained; The Dot-Dashed
curve is a $g_{T}$ exponential fit to the JLAB $G_{M}^{\ast}$ data
of Ref. \protect\cite{kamalov1} where $g_{T}(0)/g_{A}(0)=0.663 \;
c^2/GeV$, $m_{T}=0.630 \; GeV/c^2$ is obtained; Open Square data
point is from Ref. \protect\cite{ash} ) where $G_{M}^{\ast
}(0)=3.00\pm .01$; Diamond denoted data is from Ref.
\protect\cite{stuart};
 Down-Triangle denoted data is from Ref. \protect\cite{stein};  Square
denoted data is from Ref. \protect\cite{stoler93};  Open-Circle
denoted data is from Ref. \protect\cite{mistretta};  Up-Triangle
denoted data is from Ref. \protect\cite{kamalov1}. }
\label{fig1}
\end{figure}

\begin{figure}
\includegraphics[width=6.5in]{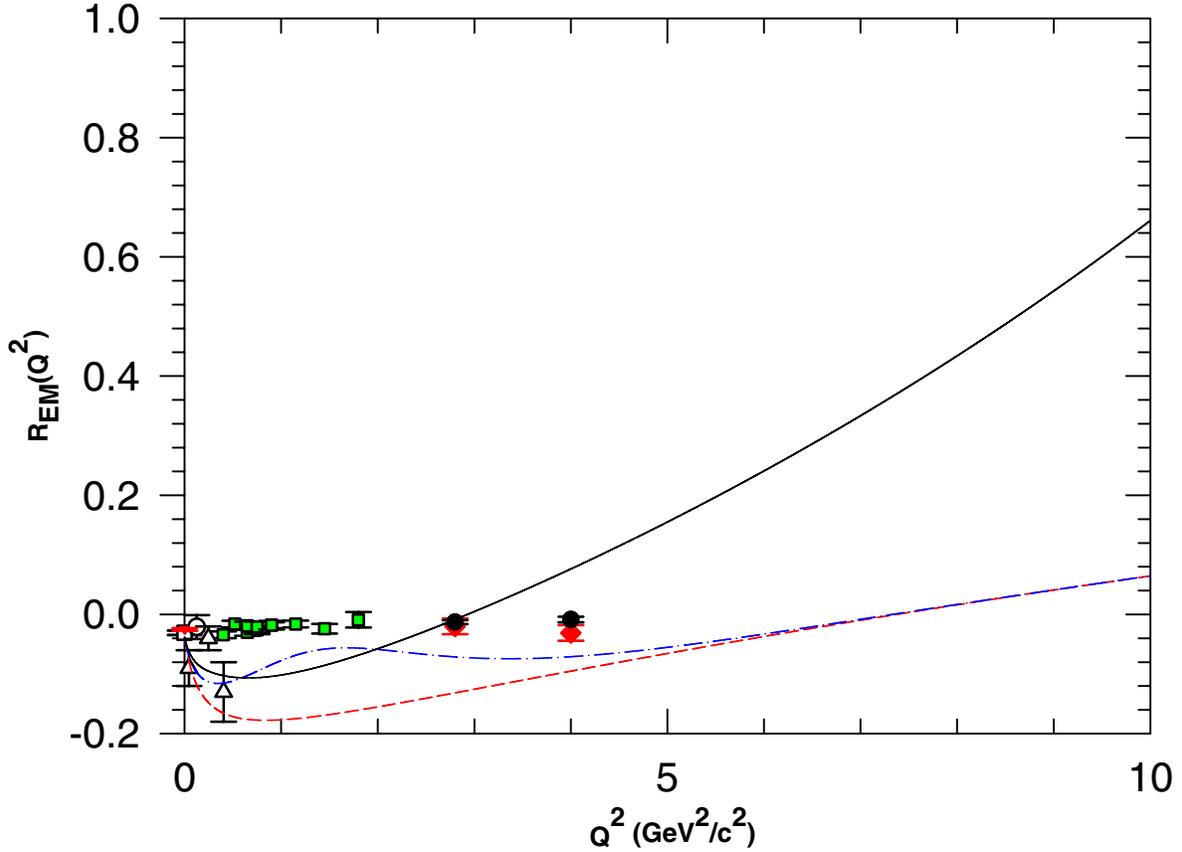}
\caption{Electromagnetic ratio $R_{EM}(Q^{2})$.  Theoretically
calculated Dashed curve with $g_{T}=0$ as discussed in the text
with $R_{EM}(0)=-0.038$; The Solid curve utilizes the results of
the $g_{T}$ dipole fit to the JLAB $G_{M}^{\ast}$ data of Ref.
\protect\cite{kamalov1} where $g_{T}(0)/g_{A}(0)=1.169 \;
c^2/GeV$, $m_{T}=0.534 \; GeV/c^2$ is obtained;  The Dot-Dashed
curve utilizes the results of the $g_{T}$ exponential fit to the
JLAB $G_{M}^{\ast}$ data of Ref. \protect\cite{kamalov1} where
$g_{T}(0)/g_{A}(0)=0.663 \; c^2/GeV$, $m_{T}=0.630 \; GeV/c^2$ is
obtained;
 Open Square data point is from Ref. \protect\cite{Blanpied};  Diamond
denoted data is from Ref. \protect\cite{frolov};  Circle denoted
data is from Ref. \protect\cite{kamalov1};  Square denoted data is
from Ref. \protect\cite{joo};  Down-Triangle data point is from
Ref. \protect\cite{beckprc61};  Open-Circle data point is from
Ref. \protect\cite{mertz};  Up-Triangle denoted data is from Ref.
\protect\cite{mistretta}. }
\label{fig2}
\end{figure}

\end{document}